# Lattice Heat Capacity of Mesoscopic Nanostructures

**B Gharekhanlou, S Khorasani, A Vafai**

School of Electrical Engineering, Sharif University of Technology,
PO Box 11365-9363, Tehran, Iran.

E-mail: khorasani@sharif.edu, vafai@sharif.edu

**Abstract.** We present a rigorous full quantum mechanical model for the lattice heat capacity of mesoscopic nanostructures in various dimensions. Model can be applied to arbitrary nanostructures with known vibrational spectrum in zero, one, two, or three dimensions. The limiting case of infinitely sized multi-dimensional materials are also found, which are in agreement with well-known results. As examples, we obtain the heat capacity of fullerenes.

## 1. Introduction

This paper presents a rigorous and unified quantization of bosonic fields, applicable to phononic, stress, strain, and electromagnetic fields. We then proceed to develop an exact relation for the lattice heat capacity of mesoscopic nanostructures, having finite physical dimensions. Extension of the model to the analysis of bulk media is shown to be agreement with known results. Numerical examples of this model for analysis of fullerenes and are shown and discussed.

## 2. Field and Energy

Suppose that a vector bosonic field state is $\langle \mathbf{r} | \mathbf{F} \rangle = \mathbf{F}(\mathbf{r},t)$, connected to a scalar energy functional as

$$\Pi = \tfrac{1}{2} \mathbf{F} : \ddot{\mathbf{K}} : \mathbf{F} = \tfrac{1}{2} \sum_{i,j=1}^{3} F_i K_{ij} F_j \tag{1}$$

where $\ddot{\mathbf{K}} = [K_{ij}]$ is a second-rank tensor, describing the inherent properties of the medium. For a lossless medium $\ddot{\mathbf{K}}$ has to Hermitian $\ddot{\mathbf{K}} = \ddot{\mathbf{K}}^\dagger = \ddot{\mathbf{K}}^t$ and real-valued. Hence, isotropic media require $\ddot{\mathbf{K}} = K\ddot{I}$, which results in $\Pi = \tfrac{1}{2} K \mathbf{F} \cdot \mathbf{F} = \tfrac{1}{2} K |\mathbf{F}|^2$. Now, using Fourier transformation pair we get

$$\mathbf{F}(\mathbf{r},t) = \frac{1}{\sqrt{2\pi}} \int_{-\infty}^{+\infty} \mathbf{B}(\mathbf{r},\omega) e^{-j\omega t} d\omega, \quad \mathbf{B}(\mathbf{r},\omega) = \frac{1}{\sqrt{2\pi}} \int_{-\infty}^{+\infty} \mathbf{F}(\mathbf{r},t) e^{-j\omega t} dt \tag{2}$$

As $\mathbf{F}(\mathbf{r},t)$ is a real field we get $\mathbf{B}(\mathbf{r},-\omega) = \mathbf{B}^*(\mathbf{r},+\omega)$, and hence the energy density expands as
$\Pi(\mathbf{r},t) = \Pi_1(\mathbf{r},t) + \Pi_2(\mathbf{r},t) + \Pi_3(\mathbf{r},t) + \Pi_4(\mathbf{r},t) =$

$$\frac{1}{4\pi} \left\{ \int_0^{+\infty}\int_0^{+\infty} \mathbf{B}(\mathbf{r},\omega) : \ddot{\mathbf{K}} : \mathbf{B}(\mathbf{r},\varpi) e^{-j(\omega+\varpi)t} d\omega d\varpi + \int_0^{+\infty}\int_0^{+\infty} \mathbf{B}(\mathbf{r},\omega) : \ddot{\mathbf{K}} : \mathbf{B}^*(\mathbf{r},\varpi) e^{-j(\omega-\varpi)t} d\omega d\varpi \right.$$

$$\left. + \int_0^{+\infty}\int_0^{+\infty} \mathbf{B}^*(\mathbf{r},\omega) : \ddot{\mathbf{K}} : \mathbf{B}(\mathbf{r},\varpi) e^{+j(\omega-\varpi)t} d\omega d\varpi + \int_0^{+\infty}\int_0^{+\infty} \mathbf{B}^*(\mathbf{r},\omega) : \ddot{\mathbf{K}} : \mathbf{B}^*(\mathbf{r},\varpi) e^{+j(\omega+\varpi)t} d\omega d\varpi \right\}$$

$$\tag{3}$$

Now if the medium allows orthogonal eigenstates $\mathbf{M}_{(n)}(\mathbf{r},t) = e^{-j\omega_{(n)}t}\mathbf{M}_{(n)}(\mathbf{r}) = e^{-j\omega_{(n)}t}\langle\mathbf{r}|(n)\rangle$ with $\langle(n)|(m)\rangle = e^{-j(\omega_{(m)}-\omega_{(n)})t}\delta_{(n)(m)} = \delta_{(n)(m)}$, and if the eigenkets $|(n)\rangle$ are complete, we get $|\mathbf{F}\rangle = \sum_{(n)} f_{(n)}|(n)\rangle$ for every bosonic field $|\mathbf{F}\rangle$ with $f_{(n)} = \langle(n)|\mathbf{F}\rangle$. Then

$$\mathbf{B}(\mathbf{r},\omega) = \frac{1}{\sqrt{2\pi}}\int_{-\infty}^{+\infty}\sum_{(n)}f_{(n)}e^{-j\omega_{(n)}t}\mathbf{M}_{(n)}(\mathbf{r})e^{+j\omega t}dt = \sqrt{2\pi}\sum_{(n)}f_{(n)}\mathbf{M}_{(n)}(\mathbf{r})\delta(\omega-\omega_{(n)}) \quad (4)$$

Hence, using (4) in (3)

$$\Pi_1(\mathbf{r},t) = \tfrac{1}{2}\sum_{(n)}f_{(n)}^2\mathbf{M}_{(n)}(\mathbf{r}):\ddot{\mathbf{K}}:\mathbf{M}_{(n)}(\mathbf{r})\exp(-j2\omega_{(n)}t) = \Pi_4^*(\mathbf{r},t) \quad (5.1)$$

$$\Pi_2(\mathbf{r},t) = \tfrac{1}{2}\sum_{(n)}|f_{(n)}|^2\mathbf{M}_{(n)}(\mathbf{r}):\ddot{\mathbf{K}}:\mathbf{M}_{(n)}^*(\mathbf{r}) = \Pi_2(\mathbf{r}) = \Pi_3^*(\mathbf{r}) \quad (5.2)$$

Integrating from (5), the total energy is obtained and given by

$$E(t) = \iiint \Pi(\mathbf{r},t)d^3r = E_1(t) + E_2(t) + E_3(t) + E_4(t) \quad (6.1)$$

$$E_2(t) = E_3(t) = \tfrac{1}{2}\sum_{(n)}|f_{(n)}|^2, \quad E_1(t) = E_4^*(t) = \tfrac{1}{2}\sum_{(n)}f_{(n)}^2 e^{-j2\omega_{(n)}t}\iiint\mathbf{M}_{(n)}(\mathbf{r}):\ddot{\mathbf{K}}:\mathbf{M}_{(n)}(\mathbf{r})d^3r \quad (6.2)$$

Here, the second and the third terms are time independent while the first and the latter ones vibrate with $\pm 2\omega_{(n)}$ frequency, and hence have zero contribution to the slowly time-varying components. Now, using (6) it is clear that the total time-average energy of the system is simply given by $E = \sum_{(n)}|f_{(n)}|^2 = \sum_{(n)}\tfrac{1}{2}(f_{(n)}^* f_{(n)} + f_{(n)} f_{(n)}^*)$, which shows that the total time-average energy of the system is equal to the sum of squared amplitudes of its eigen-modes.

*2.1. Classical Hamiltonian*

As bosons can occupy same level of energy with population more than one, this can produce a macroscopic quantum field $\mathbf{F}(\mathbf{r},t)$. We can define the functions $a_{(n)}(t) = (\hbar\omega_{(n)})^{-\frac{1}{2}}f_{(n)}e^{-j\omega_{(n)}t}$ which satisfy $\frac{d}{dt}a_{(n)}(t) = -j\omega_{(n)}a_{(n)}(t)$. This allows us to rewrite the total energy in the different form of

$$E = \sum_{(n)}E_{(n)} = \sum_{(n)}\tfrac{1}{2}\hbar\omega_{(n)}\left[a_{(n)}^*(t)a_{(n)}(t) + a_{(n)}(t)a_{(n)}^*(t)\right] \quad (7)$$

For a stress field due to phonons, a tensor representation is needed instead of vector field $\mathbf{F}(\mathbf{r},t)$, such as $\boldsymbol{\sigma}(\mathbf{r},t) = [\sigma_{ij}(\mathbf{r},t)]$ [1]. The energy density will be given by $\Pi = \tfrac{1}{2}\boldsymbol{\sigma}:\ddot{\mathbf{c}}:\boldsymbol{\sigma} = \tfrac{1}{2}\sum_{i,j,k,l}\sigma_{ij}c_{ijkl}\sigma_{kl}$, as well. Here, instead of $\ddot{\mathbf{K}} = [K_{ij}]$ which is a second-rank tensor, we have to use the fourth-rank elasticity tensor $\ddot{\mathbf{c}} = [c_{ijkl}]$, to get scalar energy density. The rest of the analysis will be exactly the same.

**3. Transition to Quantum Mechanics**

Similarity of (7) to that of a quantized harmonic oscillator suggests using a sum on the energies of oscillators with frequencies $\omega_{(n)}$, to construct the quantum mechanical Hamiltonian. So we define the ladder operators in a standard way as $\hat{a}_{(n)}(t) = \hat{a}_{(n)}(0)e^{-j\omega_{(n)}t}$ and $\hat{a}_{(n)}^\dagger(t) = \hat{a}_{(n)}^\dagger(0)e^{+j\omega_{(n)}t}$, with the algebra $[\hat{a}_n, \hat{a}_m^\dagger] = \delta_{nm}$ and $[\hat{a}_n, \hat{a}_m] = 0$. This would guarantee the independence of bosonic modes. Therefore, the quantum mechanical Hamiltonian will be

$$\mathbb{H} = \sum_{(n)}\mathbb{H}_{(n)} = \sum_{(n)}\frac{1}{2}\hbar\omega_{(n)}\left[\hat{a}_{(n)}^\dagger(t)\hat{a}_{(n)}(t) + \hat{a}_{(n)}(t)\hat{a}_{(n)}^\dagger(t)\right] \quad (8)$$

Furthermore, the ladder operators satisfy $\frac{d}{dt}\hat{a}_{(n)} = +\frac{j}{\hbar}[\mathbb{H}, \hat{a}_{(n)}]$ and likewise $\frac{d}{dt}\hat{a}_{(n)}^\dagger = +\frac{j}{\hbar}[\mathbb{H}, \hat{a}_{(n)}^\dagger]$.

## 4. Heat Capacity

Heat capacity is defined as

$$C = \partial E / \partial T \qquad (9)$$

To obtain $C$, it is enough to know the number distribution of bosons at the given temperature $T$ under thermodynamic equilibrium. Now, a phononic field under thermal equilibrium is given by [2]

$$|\psi\rangle = \sum_{\{m\}} \psi_{\{m\}} |\{m\}\rangle \qquad (10)$$

Here $\psi_{\{m\}}$ are coefficients of the expansion and $|\{m\}\rangle = |m_{(0)}\rangle |m_{(1)}\rangle |m_{(2)}\rangle \cdots |m_{(n)}\rangle \cdots$ are eigenkets, meaning that the mode $(n)$ can have $m_{(n)}$ bosons each having an energy of $\hbar\omega_{(n)}$. Now, application of Bose-Einstein statistics, requires that the probability of occupation of state $\hbar\omega_{(n)}$ with $m_{(n)}$ bosons is $P(m_{(n)}) \propto e^{-m_{(n)}\beta_{(n)}}$ with $\beta_{(n)} = \hbar\omega_{(n)}/kT$, in which $k$ is the Boltzmann's constant. So, we get

$$|\psi\rangle = \sum_{\{m\}} \sqrt{P_{\{m\}}} |\{m\}\rangle = \sqrt{N} \sum_{\{m\}} \exp\left[-\tfrac{1}{2}\sum_n m_{(n)}\beta_{(n)}\right] |\{m\}\rangle \qquad (11.1)$$

$$N = \prod_{(n)} N_{(n)} = \prod_{(n)} \left[1 - \exp(-\beta_{(n)})\right] \qquad (11.2)$$

where $N$ is the normalization constant to support $\langle\psi|\psi\rangle = 1$. A simple check shows that the Bose-Einstein distribution for bosons is readily obtained through this method as $f(E) = \left[\exp\left(\frac{E}{kT}\right) - 1\right]^{-1}$. Now, the energy of the system under the thermodynamic equilibrium at temperature $T$ is

$$E = \langle\psi|\mathbb{H}|\psi\rangle = N \sum_{\{m\}} \exp\left[-\sum_n m_{(n)}\beta_{(n)}\right] \langle\{m\}|\mathbb{H}|\{m\}\rangle \qquad (12)$$

which can be simplified through lengthy but straightforward calculations [3] as

$$E = \sum_{(n)} \hbar\omega_{(n)} \left[\exp(\beta_{(n)}) - 1\right]^{-1} + E_0 \qquad (13)$$

where $E_0 = \tfrac{1}{2}\sum_{(n)} \hbar\omega_{(n)}$ is zero point energy. Using (9) we get the exact relation for heat capacity [3]

$$C = \frac{1}{kT^2} \sum_{(n)} \left\{\hbar\omega_{(n)} \left[\exp\left(\hbar\omega_{(n)}/kT\right) - 1\right]\right\}^2 \qquad (14)$$

Small size of a nanostructure makes the energy spectrum both discrete and finite, so that (14) has a finite number of terms. This makes it possible to calculate the accurate value of (14) through numerical methods. At high temperatures, to a good approximation we have $\exp(\hbar\omega_{(n)}/kT) - 1 \approx \hbar\omega_{(n)}/kT \ll 1, \forall (n) \in \mathcal{X}^3$, so that $E \approx \sum_{(n)} kT + E_0 = LkT + E_0$, where $L$ is number of modes of the system. Thus $C \approx Lk$, will be independent of $T$. On the other hand, at very low temperatures we have $\exp(-\hbar\omega_{(n)}/kT) \approx 0, \forall (n) \in \mathcal{X}^3$, so that we get $C \approx (kT)^{-2} \sum_{(n)} [\hbar\omega_{(n)}]^2 \exp(-2\hbar\omega_{(n)}/kT)$. Therefore at the limit of zero temperature we always have $C = 0$. Furthermore, when the size of structure approaches infinity, the relation for heat capacity reduces to

$$C = \left(\tfrac{\hbar}{kT}\right)^2 \int_0^\infty \left[\omega \exp(\hbar\omega/kT) - 1\right]^{-2} D(\omega) d\omega \qquad (15)$$

with $D(\omega) = \sum_{(n)} \delta[\omega - \omega_{(n)}]$ being the density of states. But for an $N$-dimensional bulk medium, we have $D(\omega) = \alpha\omega^{N-1}$, and the corresponding lattice heat capacity will be given by direct integration as $C \propto T^N$, being in agreement with known results.

## 5. Examples

*5.1. Fullerenes: C60, C70, C80*

Fullerenes are zero-dimensional (0D) nanostructures of Carbon (figures 1,2), and their calculated heat capacities are calculated and plotted in figure 3. It is interesting to note that all heat capacities detach from the origin at $C(T=0)=0$, and furthermore they behave as $C \propto T^0 = \text{const}$ for temperatures roughly below $30\,\text{K}$.

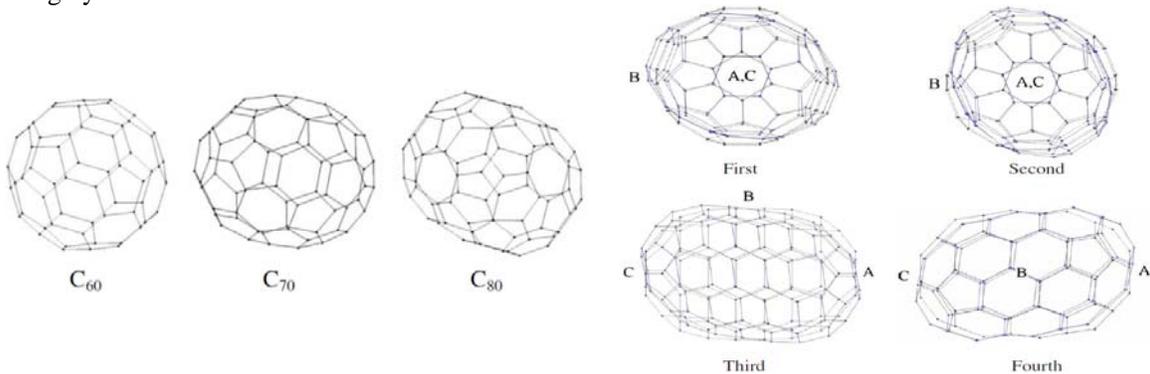

**Figure 1.** Fullerenes [4]  **Figure 2.** Vibrational modes of C80 [4]

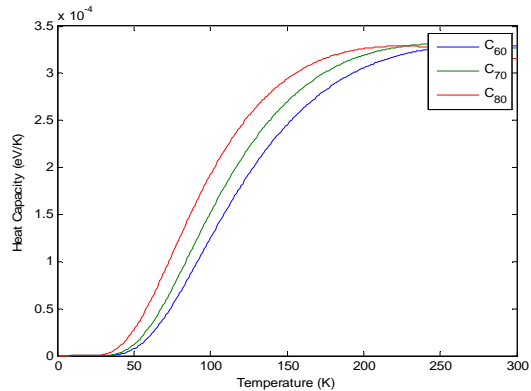

**Figure 3.** Calculated heat capacities for fullerenes.

## 6. Conclusions

We presented a rigorous and general method for quantization of bosonic fields, which was applicable to phononic fields as well. This enabled us to find the exact expression for heat capacity. We studied its behavior at high and low temperatures and also at the limit of infinite multi-dimensional structures. Agreement to well-known expressions was noticed. As application examples, we have calculated and plotted the heat capacities for C60, C70, and C80 fullerenes using their recently published spectrum.